\documentclass{elsart}
\usepackage{natbib}

\usepackage{graphics}
 \usepackage{graphicx}
 \usepackage{epsfig}
\usepackage{amssymb}

\begin{document}

\begin{frontmatter}


\title{Signatures of the Baryon Acoustic Oscillations on the Convergence
Power Spectrum of Weak lensing by Large Scale Structure}

\author[1,2]{Tong-Jie Zhang\corauthref{cor1}}\ead{tjzhang@bnu.edu.cn}
\corauth[cor1]{Corresponding author.}
\author[3]{Qiang Yuan}
\author[1]{Tiang Lan}

\address[1]{Department of Astronomy, Beijing Normal
University, Beijing, 100875, P.R.China}
\address[2]{Kavli Institute for Theoretical Physics China,
Institute of Theoretical Physics, Chinese Academy of Sciences
(KITPC/ITP-CAS), P.O.Box 2735, Beijing 100080, P.R. China}
\address[3]{Key Laboratory of Particle Astrophysics, Institute
of High Energy Physics, Chinese Academy of Sciences, P.O.Box 918-3,
Beijing 100049, P.R.China}

\begin{abstract}
We employ an analytical approach to investigate the signatures of 
Baryon Acoustic Oscillations(BAOs) on the convergence power spectrum 
of weak lensing by large scale structure. It is shown that the BAOs 
wiggles can be found in both of the linear and nonlinear convergence 
power spectra of weak lensing at about $40\le l\le600$, but they are 
weaker than that of matter power spectrum. Although the statistical 
error for LSST are greatly smaller than that of CFHT and SNAP survey 
especially at about $30<l<300$, they are still larger than their 
maximum variations of BAOs wiggles. Thus, the detection of BAOs with 
the ongoing and upcoming surveys such as LSST, CFHT and SNAP survey 
confront a technical challenge. 

\end{abstract}


\end{frontmatter}

\section{Introduction}

In the early universe prior to recombination, the free electrons couple 
the baryons to the photons through Coulomb and Compton interactions, so 
these three species move together as a single fluid. The primordial cosmological 
perturbations on small scales excite sound waves in this relativistic plasma, 
which results in the pressure-induced oscillations and acoustic peak 
\citep{1984ApJ...285L..45B,1998ApJ...496..605E}. 
The memory of these baryon acoustic oscillations (BAOs) still 
remain after the epoch of recombination. The BAOs leave their imprints 
through the propagating of photons on the last scattering surface and 
produce a harmonic series of maxima and 
minima in the anisotropy power spectrum of the cosmic microwave background 
(CMB) at $z\approx1000$. In addition, due to the significant fraction of 
baryons in the universe, BAOs can also be imprinted onto the late-time 
power spectrum of the non-relativistic matter
\citep{1984ApJ...285L..45B,1996ApJ...471..542H,1998ApJ...496..605E}, which 
have been detected in the large-scale correlation function of Sloan Digital
Sky Survey (SDSS) luminous red galaxies \citep{2005ApJ...633..560E}, and 
the power spectrum of $21$ cm emission generated from the neutral 
hydrogen from the epoch of reionization through the underlying density 
perturbation \citep{2008ApJ...673L.107M,2008PhRvL.100i1303C}. 
Essentially, the BAOs can give rise to the wiggles in the matter 
power spectrum of large scale structure during the evolution of the universe. 
Gravitational lensing can directly reveal the strenth of 
gravitational clustering\citep{2003ApJ...592..664P,2005ApJ...629...23C}, and 
weak gravitational lensing is the direct measurement of the projected 
mass distribution of the large-scale structure\citep{1999ARA&A..37..127M,
2001PhR...340..291B,2003ARA&A..41..645R,2006PhR...429....1L,2008PhR...462...67M}. 
Therefore, the BAOs prior 
to recombination should also be imprinted onto weak lensing power spectrum. 
Recently, the influence of baryons on the weak lensing power spectrum 
are investigated by many works 
\citep{2004APh....22..211W,2004ApJ...616L..75Z,2006ApJ...640L.119J}.           
%In these works, they demonstrate the change of lensing power 
%spectrum through taking into account the baryons including cooling/cooled baryons, hot baryons and star formation. 
\cite{2008arXiv0802.2416Z} study self calibration of galaxy bias in spectroscopic redshift surveys of baryon acoustic oscillations to show that SKA is able to detect BAO in the velocity power spectrum, 
and the precision measurement of cosmic magnification
are also demonstrated\citep{2005PhRvL..95x1302Z,2006MNRAS.367..169Z}.
%and do not consider the redshift distribution of lensed source galaxies 
%but assuming all lensed source are at a single redshift.
The ongoing and upcoming surveys such as the Canada-France-Hawaii-Telescope 
(CFHT) Legacy Survey\footnote{{\tt http://www.cfht.hawaii.edu/Science/CFHLS/}},
%\emph{CFHTLS}\footnote{see http://www.cfht.hawaii.edu/Science/CFHLS}, 
%\emph{SNAP}\footnote{\url{see http://snap.lbl.gov}} 
the SuperNova Acceleration Probe\footnote{{\tt http://snap.lbl.gov}: SNAP is
being proposed as part of the Joint Dark Energy Mission (JDEM)} (SNAP) 
and the Large Synoptic Survey Telescope\footnote{{\tt http://www.lsst.org}} 
(LSST) will significantly reduce the statistical errors to a few percent level 
%the sub-1$\%$ at about $l\sim100$ ($\log_{10}l\sim2$) 
in the measurement of weak lensing 
power spectrum. Therefore, it is necessary to explore the feasibility of 
detecting the BAOs in weak lensing surveys. In this paper, we concentrate on 
the wiggles of BAOs on the power spectrum of 
weak lensing by large scale structure and their detectability for current weak lensing survey. 
%We firstly calculate the matter power spectrum at different redshifts, 
%and then obtain the convergence power spectrum in the presence of baryon contents. 
%We find the wiggles in the 
%convergence power spectrum of weak lensing, and further explore the 
%detectability using the present and forthcoming weak lensing surveys.

\section{Matter Power Spectra}
We express the linear matter power spectrum in dimensionless form as 
the variance per unit logarithmic interval in wavenumber ($\log_{10} k $) 

\begin{equation}
\Delta^2(k, z)={k^3 P(k, z)\over 2\pi^2}=\delta_H^2\left 
(k\over H_0\right )^{3+n_s}\!\! T^2(k)\, {D^2(z)\over D^2(0)}
\label{pk}
\end{equation} 
where $T(k)$ and $D(z)$ are transfer function and linear growth factor 
respectively. We consider the effect of baryon content (thereby BAOs), 
so the transfer function $T(k)$ can be approximately separated into the 
cold dark matter (CDM) and baryon components: 
$T(k)=(\Omega_b/\Omega_m)T_b(k)+(\Omega_c/\Omega_m)T_c(k)$, where
$\Omega_c$ is the CDM density parameter at present, $\Omega_c=\Omega_m-
\Omega_b$. We adopt the asymptotic solutions to both $T_b(k)$ and $T_c(k)$ 
near the sound horizon given by \cite{1998ApJ...496..605E}. For convenience, 
we employ the widely used fitting formula, which are calibrated with 
N-body simulation and given by \cite{1996MNRAS.280L..19P}, 
%(hereafter PD96)
to map the linear matter power spectrum to the nonlinear power spectrum.
%Here we do not include the contribution from massive neutrinos, 
%which will be given in subsequent work of ours. 
Throughout this paper, we employ a concordance cosmological model 
revealed by the WMAP five-year observations\citep{2008arXiv0803.0732H}:
$\Omega_m=0.258$, $\Omega_\Lambda=0.742$, $\Omega_b=0.044$,
$h=0.719$, $n_s=1$.

\begin{figure}
\begin{center}
\includegraphics[angle=270, scale=0.56]{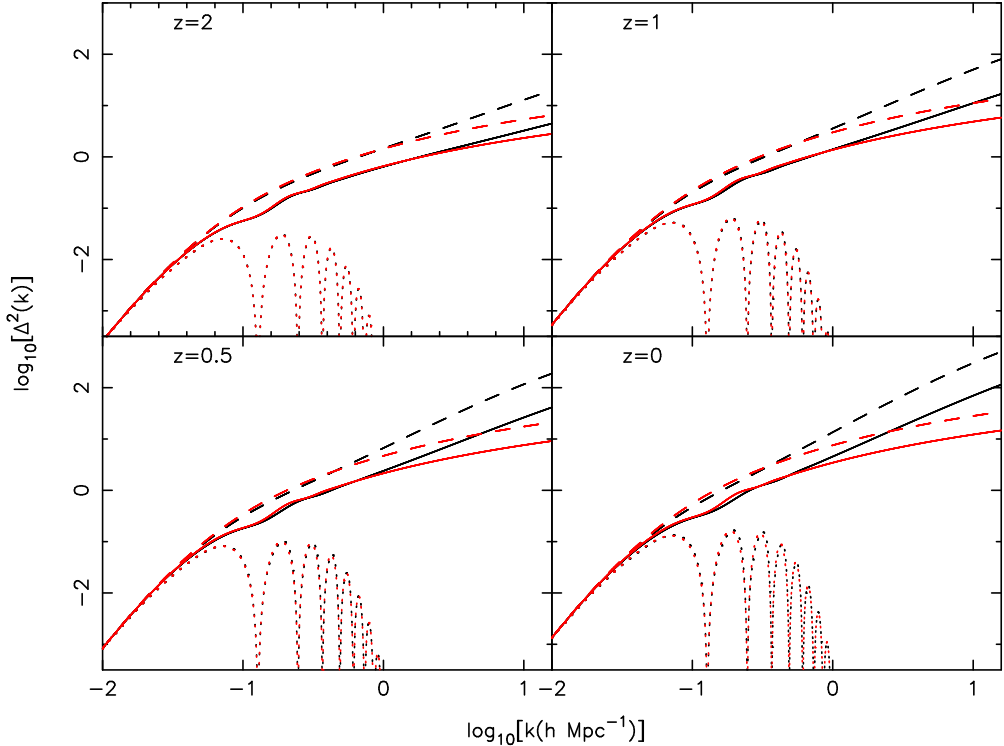}
\caption{Matter power spectra for three different matter contents:
pure CDM (dashed line), pure baryons (dotted line) and mixed
baryons+CDM (solid line) at redshift $z$=2, 1, 0.5 and 0 respectively. The
black lines correspond to the nonlinear power spectra of matter,
while the red ones the linear power spectra.}
%We find the wiggles in the convergence power spectrum of weak lensing, and 
%further explore the detectability using these weak lensing survey.
\label{fig1}
\end{center}
\end{figure}

\begin{figure}
\begin{center}
\includegraphics[angle=270, scale=0.56]{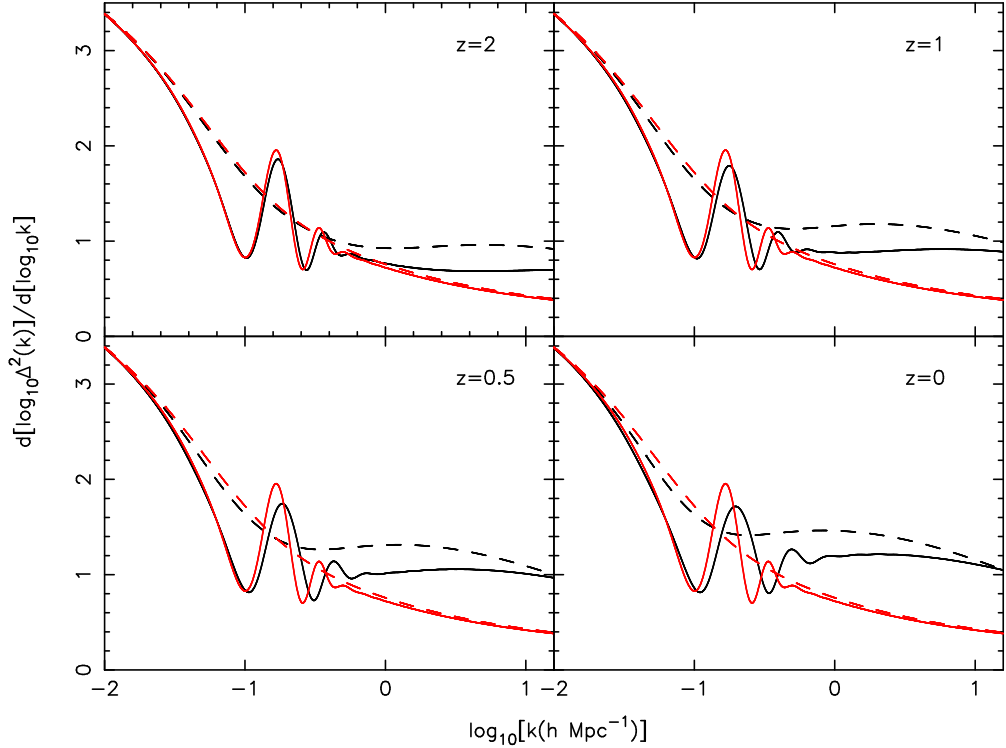}
\caption{The first derivative of power spectrum $\log_{10} \Delta^2(k)$
with respect to $\log_{10} k$ for pure CDM (dashed line) and mixed
baryons+CDM (solid line) at redshift $z$=2, 1, 0.5 and 0 respectively. The
black lines correspond to the nonlinear power spectra of matter,
while the red ones the linear power spectra. 
%Due to apparent wiggles for pure baryons, the corresponding ones are not plotted here.
}
\label{fig2}
\end{center}
\end{figure}

\begin{figure}
\begin{center}
\includegraphics[angle=270, scale=0.56]{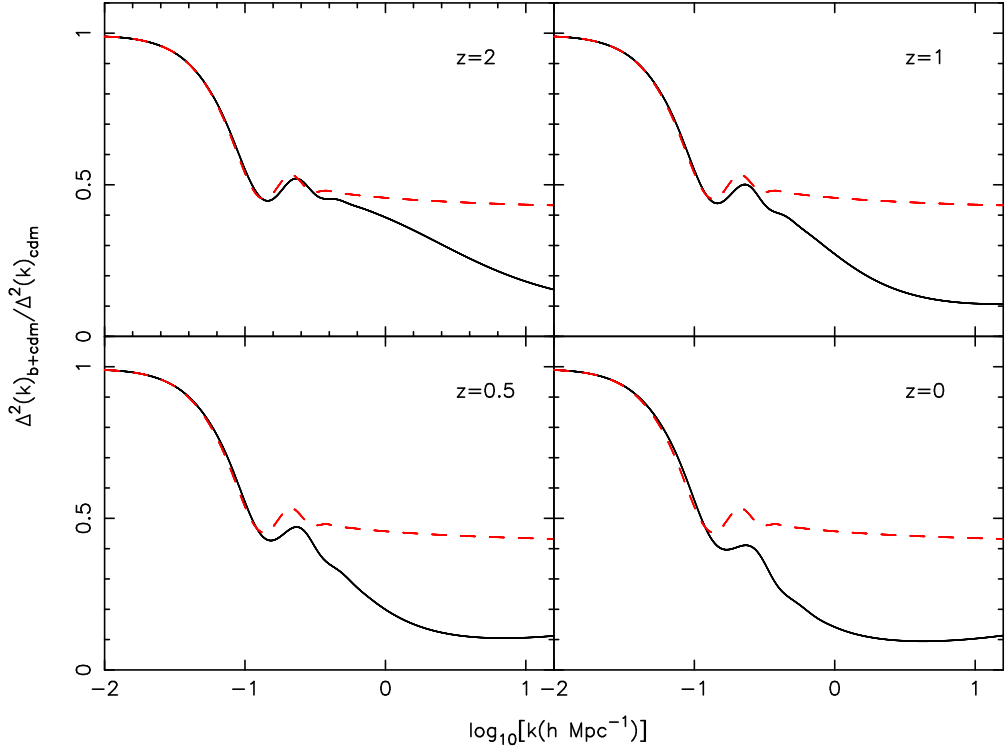}
\caption{The ratio of power spectrum for the mixed baryons+CDM to 
that for pure CDM at redshift $z$=2, 1, 0.5 and 0 respectively. The solid and 
dashed lines correspond to the ratioes of nonlinear and linear power 
spectra respectively.}
\label{fig3}
\end{center}
\end{figure}

\cite{2008ApJ...673L.107M} revealed that the signatures of BAOs, i.e., wiggles, 
are seen at wavenumber $k\sim 0.1\,h$Mpc$^{-1}$ 
%($\log_{10}k\sim-1$) 
in both of the linear and nonlinear matter power spectra with the inclusion 
of baryons at relatively high redshift $z$=6 when cosmic reionization ends. 
Due to different purpose, we in Fig.\ref{fig1} plot
the matter power spectra for three different matter contents:
pure CDM (dashed line), pure baryons (dotted line) and mixed
baryons+CDM (solid line) but at redshift $z$=2, 1, 0.5 and 0 respectively. The
black lines correspond to the nonlinear power spectra of matter,
while the red ones are the linear power spectra. In both of the linear 
and nonlinear power spectra of mixed baryons+CDM, the BAOs wiggles are 
clearly seen at wavenumber $k\sim 0.1\,h$Mpc$^{-1}$. 
%($\log_{10}k\sim-1$). 
It is also shown that, with the decrease of redshift, the clustering of structures are enhanced 
for both of pure CDM and mixed baryons+CDM models. Compared with 
pure CDM case, the mixed baryons+CDM case can suppress the linear and nonlinear 
power spectrum by an order of a few percents or more at about $k\ge 0.1\,h$Mpc$^{-1}$.
%$k\ge10\,h$Mpc$^{-1}$($\log_{10} k\ge1$). 
For clarity, we demonstrate the first derivative of 
power spectrum $\log_{10} \Delta^2(k)$ with respect to $\log_{10} k$ for 
pure CDM (dashed line) and mixed baryons+CDM (solid line) at redshift $z$=2, 1, 0.5 
and 0 respectively in Fig.\ref{fig2}. The minimum here correspond to the 
%``maximum'' 
maximum in Fig.\ref{fig1}, and about three wiggles are 
clearly revealed with the inclusion of baryon contents. We also see that, 
compared with linear power spectrum (red solid line), the nonlinear 
evolution of structure (black solid line) can suppress the amplitude of 
BAOs wiggles and shift them to small scales with decrease of redshift.   
To further show the effect of BAOs, we plot the ratio of power spectrum for 
the mixed baryons+CDM to that for pure CDM at redshift $z$=2, 1, 0.5 and 0 
respectively in Fig.\ref{fig3}. The solid and dashed lines correspond 
to the ratios of nonlinear and linear power spectra respectively.
Similar to Fig.\ref{fig2}, the wiggles for both of linear and nonlinear 
power spectrum are also seen clearly. Except for at the scales ($l$) 
around wiggles, the suppression on the amplitude of nonlinear power spectrum 
due to the inclusion of baryon contents increase with the decrease of scales, while that for linear power 
spectrum increase slightly. 
%We also notice that the relative wiggle amplitude weakly depends on the redshift. 
Thus, the BAOs signatures can be imprinted onto the entire history of cosmic 
structure evolution since the epoch of recombination. 

%\begin{figure}
%\centering
%\includegraphics[width=14cm]{fig1.eps}
%\caption{$y$-$L_{\rm x}$ relation of observational samples SZ1+Xray1
%(see Table 1). The Compton parameter $y$ are given by average over
%areas on the comving
%sizes 0.10, 0.20, 0.39, 0.78, 1.56, and 3.12 $h^{-1}$ Mpc,
%respectively. The solid lines indicate the best-fitting for all
%observational samples.}
%\end{figure}

\section{Weak Lensing Convergence Power Spectrum}
\begin{figure}
\begin{center}
\includegraphics[angle=0, scale=1.]{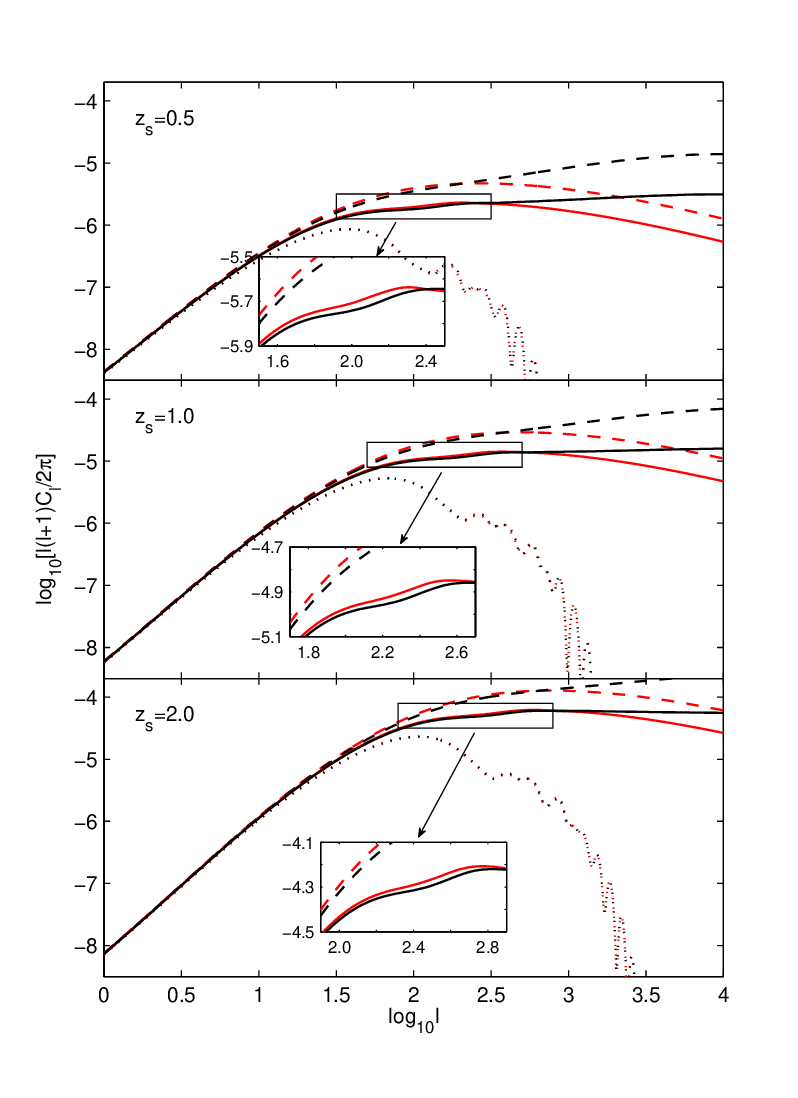}
\caption{The convergence power spectra $\log_{10}[l(l+l)C_l/2\pi]$ 
for three different matter contents: pure CDM (dashed line), pure baryons 
(dotted line) and mixed baryons+CDM (solid line) respectively. The black 
lines correspond to the nonlinear power spectra of matter, while the red 
ones are the linear power spectra. The inset shows an expanded view of 
BAOs wiggles. Here the source galaxies are distributed within a thin sheet at 
redshift $z_s=0.5,\ 1.0$ and 2.0, respectively.}
\label{fig4}
\end{center}
\end{figure}

\begin{figure}
\begin{center}
\includegraphics[angle=0, scale=0.8]{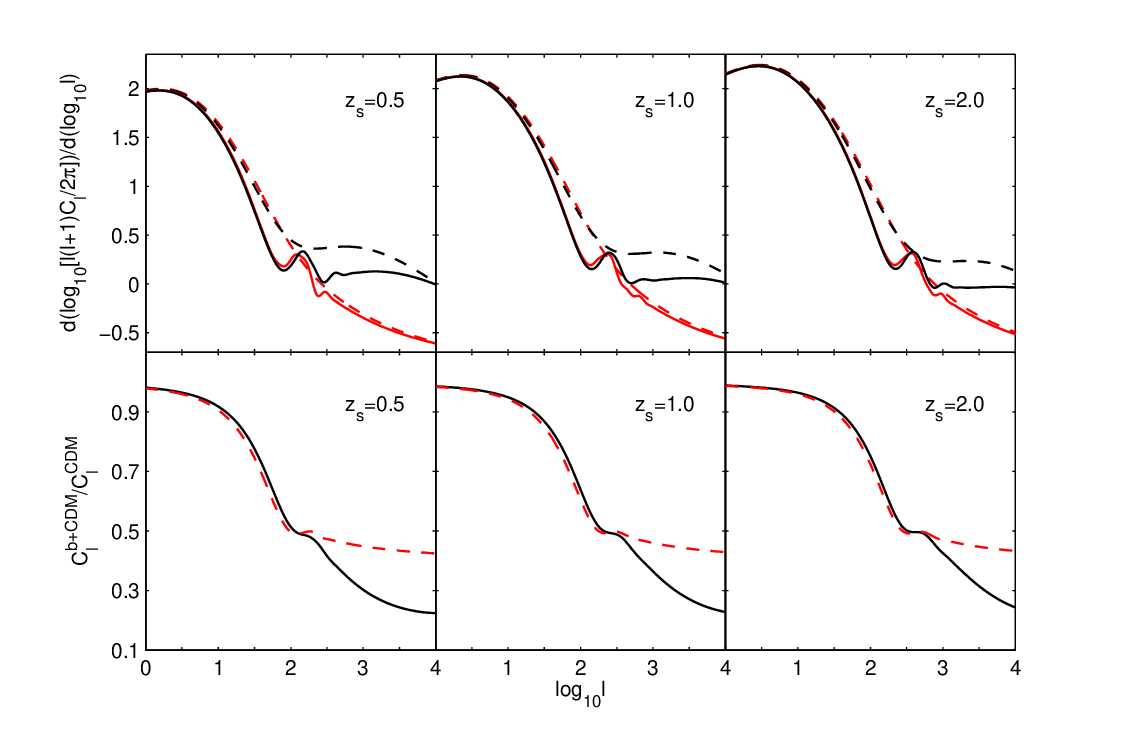}
\caption{%The source galaxies are distributed within a thin sheet at $z_s=0.5,1.0$ and 2.0, respectively. 
\emph{Upper panel} of each plot for different $z_s$: The first derivative of 
$\log_{10}[l(l+l)C_l/2\pi]$ with respect to $\log_{10}l$ for pure CDM 
(dashed line) and mixed baryons+CDM (solid line) respectively. The black 
lines correspond to the nonlinear power spectra of matter, while the 
red ones the linear power spectra. \emph{Bottom panel} of each plot for 
different $z_s$: the ratio, $C_l^{\rm{b+CDM}}/C_l^{\rm{CDM}}$ of power 
spectrum for the mixed baryons+CDM to that for pure CDM for the nonlinear 
power spectrum (black solid line) and linear power spectrum (red dashed 
line). 
%Due to apparent wiggles for pure baryons, the corresponding derivative
%and difference are not plotted in both of panels. 
}
\label{fig5}
\end{center}
\end{figure}

Through the matter power spectrum, BAOs signature enters into the 
statistics of weak lensing by large scale structure. Using Limber's 
approximation\citep{1992ApJ...388..272K,2001PhR...340..291B,2007A&A...473..711S}, 
we can write the convergence power spectrum of weak 
lensing as

\begin{equation}
C_l=\int_0^{\chi_s} d\chi \frac{W^2(\chi)}{r^2(\chi)}
P(l/r(\chi), z)
\label{eq:cl}
\end{equation}
where $r(\chi)$ is the radial comoving coordinate distance and $r(\chi)=
\sinh(\chi)$ for open, $r(\chi)=\chi$ for flat and $r(\chi)=\sin(\chi)$ 
for closed geometry of Universe respectively. The weight function $W(\chi)=\frac{3}{2}
\Omega_{m}H_0^2g(\chi)(1+z)$ is determined by the source galaxy 
distribution function $n(z)$ and the lensing geometry 
$g(\chi)=r(\chi) \int_{\chi}^{\infty} d\chi' n(\chi')\frac{r(\chi'
-\chi)}{r(\chi')}$. Here $n(z)=n(\chi)d\chi/dz$ is normalized such
that $\int_0^{ \infty} n(z)dz=1$. 
%For the CFHT Legacy Survey, we
%adopt $n(z)=\frac{\beta}{z_0\Gamma(\frac{1+\alpha}{\beta})}(\frac{z}{z_0})^
%\alpha\exp(-(\frac{z}{z_0})^\beta)$ with $\alpha=2$ and
%$\beta=1.2$ and the source redshift parameter $z_0=0.44$, which
%peaks at $z_p=1.58z_0$, respectively. The mean redshift is
%$\bar{z}=2.1z_0$ and the median redshift isbaryon contents
%$z_h=1.9z_0$\citep{2002A&A...393..369V}. 
If all sources are at a single redshift $z_s$ (or the distribution of source 
galaxies is given within a thin sheet at redshift $z_s$), we have 
$n(z)=\delta(z-z_s)$ and then $g(\chi)=r(\chi) r(\chi_s-\chi)/r(\chi_s)$. 
Following the works \citep{2008ApJ...672...19R,2006ApJ...640L.119J}, for 
simplicity, we assume that the source galaxies are distributed within a 
thin sheet at $z_s$ throughout this paper. For a flat universe, 
$d\chi=dr=dz/H(z)$, so Eq.(\ref{eq:cl}) reduces to
\begin{equation}
C_l={2\pi^2\over l^3}
\int_0^{z_s} dz \,{W^2(z) \,r(z)\over H(z)} \Delta^2(l/r(z), z).
\label{eq:cll}
\end{equation}
In Eqs.(\ref{eq:cl}) and (\ref{eq:cll}), the matter power spectrum as a 
function of wavenumber $k=l/r(z)$ and redshift $z$ satisfies Eq.(\ref{pk}): 
$\Delta^2(l/r(z),z)=k^3 P(l/r(z),z)/2\pi^2$. In Fig.\ref{fig4}, we 
show the signatures of BAOs on the convergence power spectra of weak 
lensing with the source galaxies at redshift $z_s=0.5,\ 1.0$ and 2.0, respectively. 
This figure plot the 
convergence power spectra $\log_{10}[l(l+l)C_l/2\pi]$ for three different 
matter contents: pure CDM (dashed line), pure baryons (dotted line) and 
mixed baryons+CDM (solid line) respectively. 
%The black lines correspond to the nonlinear power spectra of matter, 
%while the red ones are the linear power spectra. 
Both of the linear and nonlinear power spectra of 
weak lensing are suppressed at about $l>10$ due to the inclusion of baryon contents.
The BAOs wiggles can be visible in both of the linear and 
nonlinear power spectra for the mixed baryons+CDM model at about 
$40\le l\le600$. 
%($1.6\le\log_{10}l\le2.8$). 
With the increase of $z_s$ from 0.5 to 2, 
the wiggles are shifted to small scales for both of linear and nonlinear 
power spectra, i.e., from the scale range of about $40\le l\le250$ 
%($1.6\le\log_{10}l\le2.4$) 
to $100\le l\le600$. 
%($2.0\le\log_{10}l\le2.8$). 
Compared with that of matter power spectrum, the BAOs wiggles in the convergence 
power spectra are weaker. 
%We notice that the wiggles are slightly suppressed with the increase of source galaxies redshift $z_s$. 
To be more clearly, the first derivative of $\log_{10}[l(l+l)C_l/2\pi]$ 
with respect to $\log_{10}l$ and the ratio, $C_l^{\rm{b+CDM}}/C_l^{\rm{CDM}}$ 
of power spectrum for the mixed baryons+CDM to that for pure CDM are 
shown in Upper panel and Bottom panel of Fig.\ref{fig5} respectively. 
At about $40<l<600$, 
%($1.6<\log_{10}<2.8$),
the wiggles are revealed clearly 
in Upper panel for different $z_s$ due to the inclusion of baryon contents. 
%The suppression of amplitude of wiggles with increase of source galaxies redshift $z_s$ are also seen.  
Similar to matter power spectrum (Fig.\ref{fig3}), it is shown in 
Bottom panel for different $z_s$ that the suppression on 
nonlinear convergence power spectrum amplitude due to the inclusion of baryon contents increases with the decrease 
of scales ($l$) except for the scales around wiggles, while
that for linear convergence power spectrum vary slightly.  

\section{Detectability}
\begin{figure}
\begin{center}
\includegraphics[angle=0, scale=0.7]{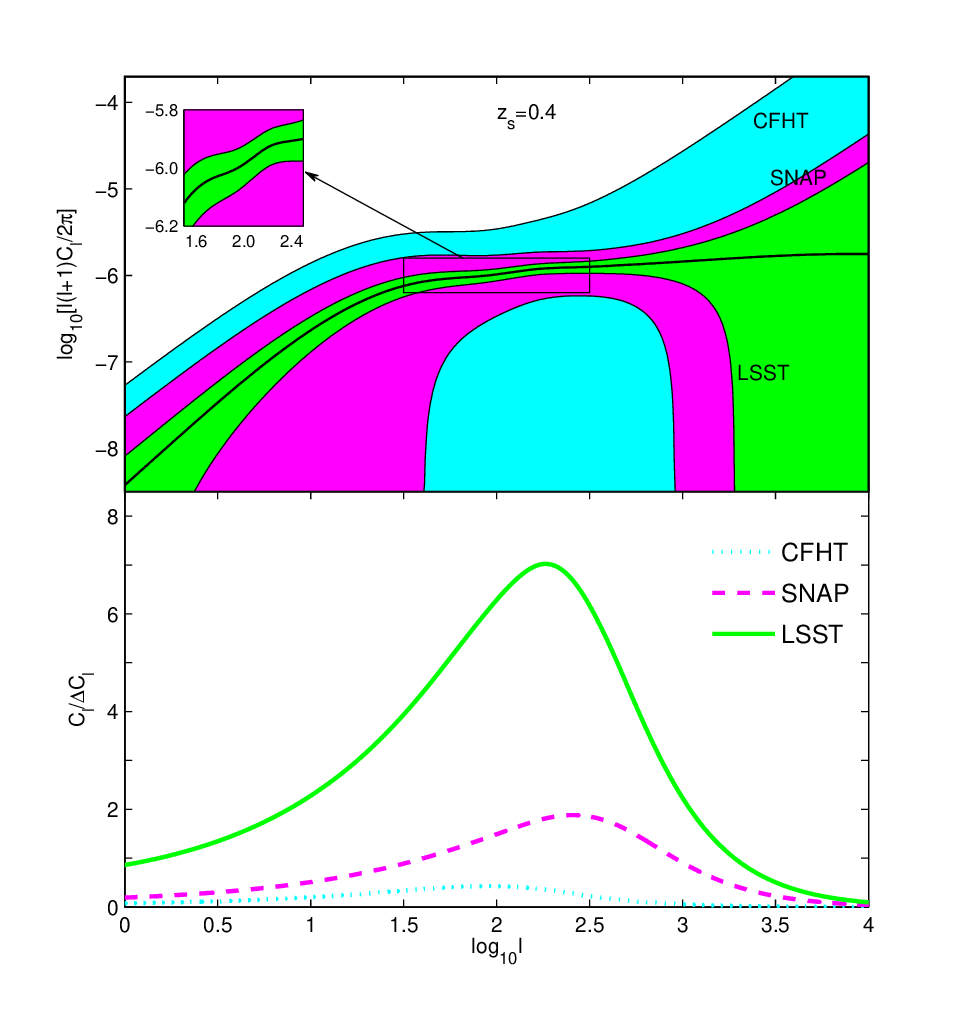}
\caption{
\emph{Top panel}: The nonlinear convergence power spectra
$\log_{10}[l(l+l)C_l/2\pi]$ just for mixed baryons+CDM with statistical
errors. The three shaded regions represent the statistical errors expected
by the CFHT (cyan), SNAP (magenta), and LSST (green) from outermost to
innermost bands, respectively.
\emph{Bottom panel}: The signal-to-noise $C_l/\Delta C_l$ of the convergence
power spectrum for the CFHT (cyan dotted line), SNAP (magenta dashed
line), and LSST (green solid line)
}
\label{fig6}
\end{center}
\end{figure}

In order to demonstrate the detectability of BAOs, we compare our results 
with the statistical uncertainty in the measurement of the convergence 
power spectrum by ongoing and upcoming weak lensing survey projects such 
as CFHT, SNAP and LSST. The statistical errors in the measurements of 
weak lensing power spectrum $C_l$ (assuming Gaussianity) are described by
\citep{1992ApJ...388..272K,1998ApJ...498...26K,1998ApJ...506...64S,2002PhRvD..65f3001H}
\begin{equation}
\Delta C_l=\sqrt{2\over(2\ell +1)f_{\mathrm{sky}}}(C_l+\frac{\gamma_{\rm 
int}^2}{\bar n_g}),
\label{delcl}
\end{equation}
where $f_{\mathrm{sky}}=\Theta^2\pi/129600$ is the coverage fraction of 
sky covered by a survey of area $\Theta^2$ in units of deg$^2$, 
$\bar n_g$ is the effective surface number density of source galaxies 
with measurable shapes on the sky, and $\gamma_{\rm int}$ is the 
{\tt rms} intrinsic shape noise for each galaxy. 
The cosmic variance, which corresponds to the first term of Eq.(\ref{delcl}),
dominates the uncertainty in the observed convergence power spectrum
on large scales, and Poisson noise corresponding to the second term
come from the limited number of galaxies on small scales.
As an ongoing 
observational project, we consider a survey CFHT which covers a 
fraction $f_{\mathrm{sky}}=4 \times 10^{-3}$ of the sky, with a number 
density of source galaxies of $\bar n_g=13 \ \mathrm{arcmin}^{-2}$. 
For future survey, we take $f_{\mathrm{sky}}=0.5$ and $\bar n_g=50 \ 
\mathrm{arcmin}^{-2}$ for LSST survey and $f_{\mathrm{sky}}=0.025$ 
and $\bar n_g=100 \ \mathrm{arcmin}^{-2}$ for SNAP survey respectively
\citep{2008ApJ...672...19R}. We adopt $\gamma_{\rm int}=0.22$ for the 
variance of the intrinsic ellipticity of source galaxies. 
%Note that the estimates of statistical uncertainties here are just under the 
%assumption of Gaussian statistics. 
%\emph{Top panel} of 
The top panel of Fig.\ref{fig6} shows the nonlinear convergence power spectra $\log_{10}[
l(l+l)C_l/2\pi]$ for mixed baryons+CDM with statistical errors with 
the source galaxies distributed within a thin sheet at $z_s=0.4$..
%The three shaded regions represent the statistical errors expected
%by the CFHT (cyan), SNAP (magenta), and LSST (yellow) from outermost 
%to innermost bands, respectively. 
For CFHT and SNAP survey, the statistical error are greatly larger than the 
%relative amplitude of wiggles. 
maximum variations of wiggles, which is the power spectrum difference between 
wiggle peak and adjacent troughs. Compared with that of CFHT and SNAP survey, the  
statistical error for LSST are relatively small especially at about 
$30<l<300$. In addition, we also in the bottom panel of
Fig.\ref{fig6} plot the signal-to-noise $C_l/\Delta C_l$ of the convergence
power spectrum for the weak lensing survey projects. It is noticed that the range of scales
with high signal-to-noise for LSST is much wider than that of CFHT and SNAP.
%($1.5<\log_{10}<2.5$). 
Even so, it seems that a significant detection of BAOs wiggles on  
convergence power spectra is still difficult for current weak lensing survey due 
to the weakness of BAOs signals themselves and limitation of statistical errors of 
current telescopes. Our result is roughly in agreement with that of work\citep{2006ApJ...647L..91S}.
%In order to further explore the 
%observational feasibility of wiggles for LSST, we plot the nonlinear convergence power spectra 
%$\log_{10}[l(l+l)C_l/2\pi]$ for mixed baryons+CDM with statistical errors 
%by extending the coverage fraction of sky from $f_{\mathrm{sky}}$=0.5 to  
%1 (whole sky) in \emph{Bottom panel} of Fig. \ref{fig6}. It is shown 
%that, even for the whole sky survey of LSST, the statistical error are 
%still larger than the maximum variations of wiggles. 

%\begin{figure}
%\begin{center}
%\includegraphics[angle=0, scale=0.6]{fig7-LSST-zs04.eps}
%\includegraphics[angle=0, scale=0.6]{fig6-4s-zs05.eps}
%\caption{Same as Fig. \ref{fig6} but just for LSST. The two shaded regions represent the statistical errors %expected by LSST with the coverage fraction of  $f_{\mathrm{sky}}$=0.5(blue) and 1(red) from outermost to %innermost band, respectively.}
%\label{fig7}
%\end{center}
%\end{figure}

\section{Conclusions}
In this paper, we use an analytical approach to present signatures of 
BAOs on the convergence power spectrum of weak lensing. 
%We firstly calculate the matter power spectrum at different redshift. 
We show that, in both of the linear and nonlinear matter power spectra of mixed 
baryons+CDM, the BAOs wiggles can be clearly seen at about 
$k\sim 0.1\,h$Mpc$^{-1}$. With the decrease of redshift, the clustering of structures are enhanced 
for both of the pure CDM and mixed baryons+CDM models. Compared with 
pure CDM case, the mixed baryons+CDM case can suppress the linear and nonlinear 
matter power spectrum by an order of a few percents or more at about $k\ge 0.1\,h$Mpc$^{-1}$.
The nonlinear evolution of structure can suppress the amplitude of 
BAOs wiggles and shift them to small scales with decrease of redshift.
For the convergence power spectrum of weak lensing, we show that 
the BAOs wiggles can be visible in both of the linear and 
nonlinear power spectra for the mixed baryons+CDM model at about 
$40\le l\le600$, but they are weaker than that of matter power spectrum. 
With the increase of $z_s$ from 0.5 to 2, the BAOs wiggles are shifted to small scales for both of linear and nonlinear 
power spectra, i.e., from the range of about $40\le l\le250$ to $100\le l\le600$.  
We also study the detectability of BAOs's wiggles for 
weak lensing survey. Although the statistical error for LSST are greatly smaller than that of CFHT and SNAP survey 
especially at about $30<l<300$, they are still larger than the their maximum variations of BAOs wiggles.
Thus, the detection of BAOs with the ongoing and upcoming surveys such as LSST, CFHT and SNAP survey i
confront a technical challenge. Therefore, we expect future weak lensing survey with more lower 
statistical errors to capture BAOs wiggles signal from the convergence power spectrum. In addition, 
we have shown that the BAOs signatures can be imprinted onto the entire history of cosmic 
structure evolution since the epoch of recombination. The future observation of BAOs on weak 
lensing will provide a complementary knowledge of current BAOs measurement. 
On the other hand, the BAOs provide a ``standard ruler'' for the determination of cosmological 
parameters especially the probes of dark energy, so 
the combination of BAOs measurement by weak lensing together 
with other detection of BAOs signature at different stage: the last scattering surface, reionization and large scale distribution of galaxies in local universe, will supply more robust constraint on cosmological models in the future.    

\noindent{\bf Acknowledgments.}
We are very grateful to the anonymous referee for his valuable
comments and suggestions that greatly improve this paper.
We also thank Dr.Pengjie Zhang for valuable discussion and useful comment.
This work was supported by the National
Science Foundation of China (Grants No.10473002, 10533010), the Ministry of
Science and Technology National Basic Science program (project 973)
under grant No.2009CB24901 and the Scientific Research Foundation for the Returned
Overseas Chinese Scholars, State Education Ministry.

%\begin{thebibliography}{}
% \bibitem[Names(Year)]{label} or \bibitem[Names(Year)Long names]{label}.
% (\harvarditem{Name}{Year}{label} is also supported.)

%\bibitem[{{Boldyrev} {et~al.}(2004){Boldyrev}, {Linde}, \&
%  {Polyakov}}]{2004PhRvL..93r4503B}
%{Boldyrev}, S., {Linde}, T., \& {Polyakov}, A. 2004, Physical Review Letters,
%  93, 184503
%\end{thebibliography}
%\begin{thebibliography}{21}

%\bibliography{ztjcos-lensbib}
%\bibliographystyle{apj}
\clearpage
\end{document}